\documentclass[twoside]{article}

\usepackage[preprint]{aistats2027}
\usepackage[T1]{fontenc}
\usepackage[utf8]{inputenc}
\usepackage{amsmath}       
\usepackage{graphicx}
\usepackage{booktabs}
\usepackage{microtype}
\usepackage{xcolor}
\usepackage{url}
\usepackage[hidelinks]{hyperref}
\usepackage{float}
\usepackage{dblfloatfix}   

\graphicspath{{figures/}{./figures/}{../generated/mock_paper_v2/assets/}{../generated/v2_boltz_function_figures/plots/}}

\newcommand{\pkd}{\ensuremath{\mathrm{p}K_{\mathrm d}}}

\begin{document}
\raggedbottom   

\runningtitle{Confident and Wrong}
\runningauthor{LoPrete et al.}

\twocolumn[
  \aistatstitle{Confident and Wrong: Structure Confidence Is a Poor Selection Filter for Protein Function}

  \aistatsauthor{Eric LoPrete, Pavle Jeremić}

  \aistatsaddress{Aether Biomachines \\ Correspondence: \texttt{eric@aetherbio.com}}
]

\begin{abstract}
Structure-guided protein design can generate plausible candidates, but structural confidence does not predict function. We trained a function-guided model on measured metal-binding affinity with protein sequence, ligand identity, and assay conditions, and used it to design rare earth element (REE) binding variants of lanmodulin, a lanthanide-binding protein with four EF-hand domains. We screened 297 designs from three strategies against terbium and dysprosium, adjacent lanthanides among the hardest REE pairs to separate.
Selection on structural confidence inverted the outcome it was meant to improve. The Structure-Confidence-Selected panel carried the highest pTM, complex ipLDDT, and composite confidence scores, and expressed at 97\%, yet only 16.7\% of its designs bound terbium (95\% CI 10.4 to 25.7), against 83.3\% (74.6 to 89.5) for the function-guided panel, whose confidence scores were the lowest. The odds of binding differed 25-fold between the two panels (Fisher exact $p = 1\times10^{-20}$). No structural-confidence metric separated binders from non-binders consistently across panels, whereas the function-trained model's predicted \pkd{} gave the highest ROC AUC in every panel, 0.79 to 0.92, and ranked measured affinity among function-guided binders at Spearman $\rho = 0.98$.
On the design objective, Tb/Dy selectivity, the function-guided designs also exceeded the natural lanmodulin family, outranking a randomly drawn ortholog 84\% of the time across 149 orthologs. Structural confidence is a plausibility signal, not a selection criterion for function; direct supervision on measured function is what ranks designs.
\end{abstract}

\section{Introduction}

\subsection{Protein Engineering for Chemical Manufacturing}

New methods for producing and separating chemicals could have profound effects
on manufacturing, supply chains, and the materials economy. Proteins offer a
route to programmable chemical engineering: their sequences can be modified,
their functions can be screened experimentally, and successful designs can be
manufactured biologically.

We combine machine learning with high-throughput
screening to optimize the function of industrially relevant proteins. The aim
is to develop proteins that can be deployed in efficient chemical production
and separation processes. More broadly, our goal is to develop a general protein function model which can design novel proteins for novel tasks, at and beyond the frontier of protein mechanism explanations.

Recent advances in machine learning have transformed protein modeling.
AlphaFold made accurate structure prediction broadly accessible
\cite{alphafold}, while RFdiffusion and related systems enabled the generation
of proteins with prescribed structural features \cite{rfdiffusion}. These
methods have expanded the set of protein structures that engineers can explore.

\subsection{Learning Protein Function from Measurements}

The promise of machine learning is not limited to structure prediction.
Protein language models such as ESM learn useful representations from large
collections of natural protein sequences \cite{esm}. Closed- and open-loop
systems can combine such representations with experimentally measured labels
to learn aspects of protein function \cite{enzymeai}.

Directed evolution follows a related logic: proteins are screened for function,
promising variants are selected, and mutations are introduced to explore new
functional regions \cite{directedevolution}. Machine-learning models can help
identify candidate mutations and can be updated over successive design--build--test--learn cycles \cite{gpfitness}. High-throughput laboratories make it
possible to generate the cohesive functional labels needed for these models.

Data remain the central constraint. Models trained by example require many
measurements that span the relevant experimental space. Structure-prediction
models benefit from large collections of crystallographic, cryogenic electron
microscopy, and NMR structures; function-prediction models analogously require
well-controlled functional measurements.

Static structures are informative but incomplete descriptions of biochemical
function. Co-folding models can assign high confidence to structures even when
the expected functional consequences of sequence changes are absent
\cite{cofoldingphysics}. Protein dynamics, solvent, ions, and assay conditions
can also contribute to the observed outcome \cite{enzymedynamics}. These
limitations motivate direct modeling of measured function alongside structural
analysis.

\subsection{A Function-Guided Design Strategy}

The central problem is therefore one of data integration. Instead of converting
a sequence into a three-dimensional structure and then deriving function from a
hand-built interpretation of that structure, we seek to predict observed
function directly from primary sequence, chemical context, and assay
conditions. The inputs include the protein sequence, buffer composition, salts,
cofactors, temperature, pressure, expression method, and other available
descriptors of the reaction.

Mapping these inputs to kinetic or equilibrium constants makes protein function
comparable across assays through quantities such as catalytic turnover or
dissociation kinetics. The resulting model can be used to define objectives for
ligand targeting, selectivity, or performance under specified conditions. This
shifts part of the design burden from an explicit residue-level structural
rationale to a model learned from experimental outcomes.

We demonstrate this approach by comparing three protein-design strategies that
require different degrees of structural hypothesis.

\section{Methods}

\subsection{Function-Model Architecture}

\begin{figure}[H]
  \centering
  \includegraphics[width=\columnwidth]{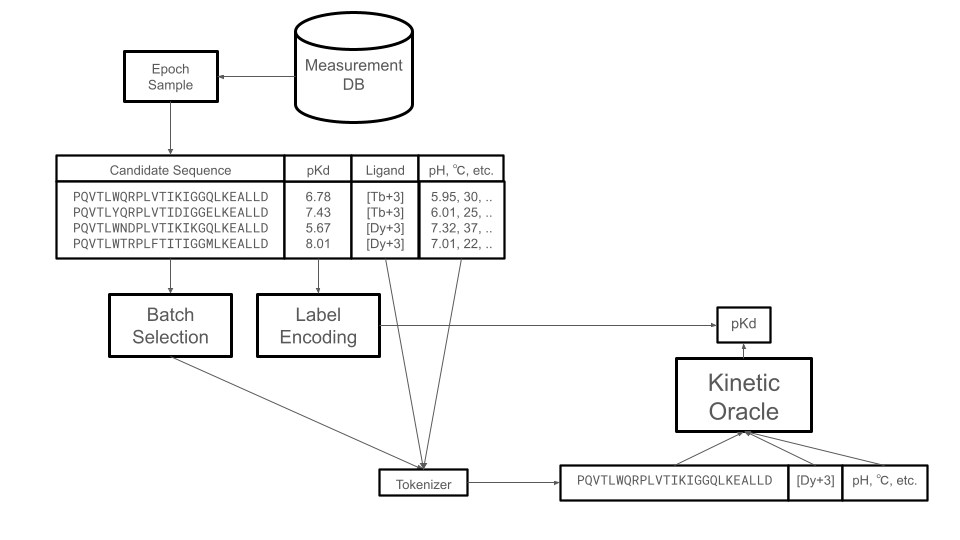}
  \caption{\textbf{Kinetic Oracle Training.} Diagram of how the function data is used to train a function oracle module.}
  \label{fig:oracle-training}
\end{figure}

The function model used a pair of transformer models. One model learns how a sequence and reaction predict a function, and the other learns how to change a sequence to make it more functional. Both leverage the same training data, but inverted to enable the different predictive capabilities. We use the protein sequence, metal-ion representation, buffer representation,
assay conditions, and the measured function
\(\pkd = -\log_{10}(K_{\mathrm d})\). All input tokens participate in
attention in both models (Figure~\ref{fig:oracle-training}).

\begin{figure}[H]
  \centering
  \includegraphics[width=\columnwidth]{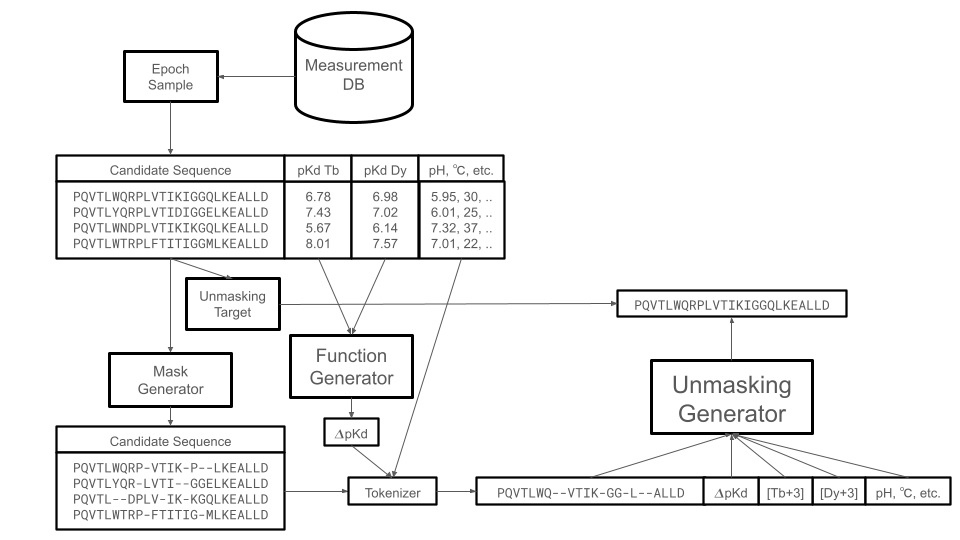}
  \caption{\textbf{Sequence Generator Training.} Diagram of how the function data is used to train a sequence generator module.}
  \label{fig:generator-training}
\end{figure}

Next is the model that determines how to mutate a given protein. To compile a label, we sample from the same measured kinetic data to get a protein observed with different metals in the same conditions. A function description is assembled from these multiple observations, and the input token sequence contains a masked representation of a protein, the different metals we want selectivity for, and the conditions. The label consists of the unmasked representation of the protein. The model is trained to unmask the protein, guided towards a protein with the queried function, such as ligand selectivity
(Figure~\ref{fig:generator-training}).

The generator and oracle can be pre-trained with unsupervised natural protein unmasking, similar to ESM \cite{esm}. Together, this gives us two different models, trained using the same measured data, but with different orientations of features and labels. The oracle takes a sequence and predicts the function, the generator takes the function and generates a sequence. These can operate cooperatively to generate new proteins.

\subsection{Design Strategies}

All three strategies began from lanmodulin, a lanthanide-binding protein with
four EF-hand domains and function adjacent to the target outcome of rare-earth
element selectivity \cite{lanmodulinextraction}. We focused on Tb and Dy, two
magnet dopants with closely related chemistry. Throughout, selectivity denotes
the fold difference between the two dissociation constants,
\(10^{|\Delta \mathrm{p}K_\mathrm{d}|}\), where
\(\Delta \mathrm{p}K_\mathrm{d} = \mathrm{p}K_\mathrm{d,Tb} - \mathrm{p}K_\mathrm{d,Dy}\).
It is a magnitude: a protein that binds both metals equally scores 1, and the
statistic does not record which of the two is bound more tightly. To keep the comparison
tractable, designs could modify EF-hand regions but not the remainder of the
backbone. The structure-guided methods used released model weights and were not
trained on our binding measurements. The function-guided model was trained on
approximately 2,500 proteins measured in-house.
The model was trained on a diverse set of metal binding proteins of different sizes and structures, but heavily biased with lanmodulin variants. The proteins in this study, lanmodulin orthologs, the Function-Selected, Co-Generation-Selected, and Structure-Confidence-Selected protein sets were not used in training this model.

\subsubsection{Function-Selected Design}

\begin{figure}[H]
  \centering
  \includegraphics[width=\columnwidth]{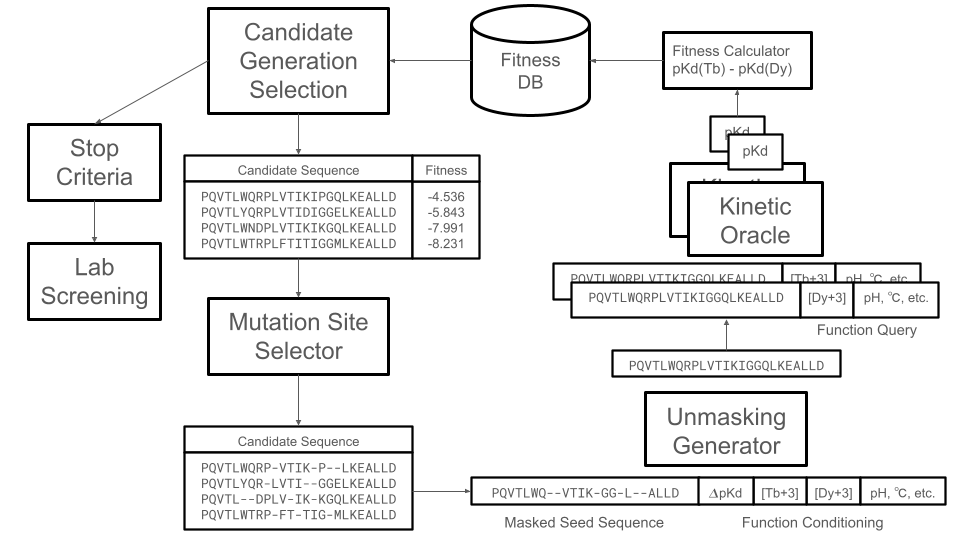}
  \caption{\textbf{Sequence Design.} Diagram of how the kinetic oracle and sequence generator are used in tandem to guide in-silico evolution of protein sequences.}
  \label{fig:sequence-design}
\end{figure}

Our function-guided strategy used the two models in a recursive
in-silico evolution loop. Pseudo-random residue sampling
determined the specific sequence positions to mask, and the generator model unmasked them with the target function as conditioning. One residue is unmasked at a time, and that partially unmasked sequence is fed back into the generator input until the sequence is completely unmasked. The generator's residue-head softmax probability predictions can be sampled stochastically to control exploration. The oracle is used to score each protein. These are used to calculate a fitness, which informs the system of which variants graduate to the next generation (Figure~\ref{fig:sequence-design}).

The fitness we used in this campaign was simply to maximize the absolute value of the difference in \pkd{} while maintaining a minimum \pkd{} to facilitate measurability.
This cycle is repeated until a stopping criterion is met, and the final top-ranked proteins are promoted to be screened by the laboratory. The stopping criterion must be defined carefully to prevent hallucinatory exploitation of the oracle, mostly by limiting the number of in-silico generation loops. This exact number should be determined experimentally. However, because sequences that exploit the oracle are still measured experimentally, the data they return can be valuable when exploration of novel motifs is the goal.

This approach has several advantages. The generator learns evolutionarily conserved residues from the wider protein universe, and fine-tuning teaches conserved residues in the function universe. Thus, a masked highly conserved residue is likely to be restored to the conserved identity, minimizing the occurrence of degenerate mutations. Conversely, less conserved residues can be exploited for functional contributions. This alleviates some burden on the protein engineers to identify which residues to modify or avoid.

\subsubsection{Structure-Confidence-Selected Design}

The next strategy used a traditional rational-design workflow. Chemical
intuition and coordination hypotheses were used to generate candidate
three-dimensional binding sites with molecular dynamics and Architector
\cite{architector}. RFdiffusion and ProteinMPNN \cite{proteinmpnn} were then
used to design candidate sequences, and structures were filtered using metrics
including pLDDT and AlphaFold RMSD. This process selected 93 candidates from
approximately 10,000 generated proteins. It required an explicit mechanistic
hypothesis at several stages and therefore carried the highest rationale burden.
Protein engineers must study 3D structures to identify binding sites, active residues, and formulate a hypothesis for how to modify them to improve function.

\subsubsection{Co-Generation-Selected Design}

The final strategy used DISCO, a single model system that co-generates sequence
and structure around a seeded ligand \cite{disco}. The target metal served as
the principal function-defining input, and the model generates a protein around it, including the active residues, reducing the amount of residue-level
mechanistic specification required at design time. For metal binding, this is mostly trivial, but for more complex ligands and substrates, carefully optimized transition-state geometry must be prepared. Generation required roughly
eight hours per 25 proteins in the evaluated workflow, yielding 108 candidates
without an additional oversampling-and-filtering stage, 104 of which were
carried forward to screening.

\subsection{Boltz-2 Structural Confidence Metrics}

Boltz-2 \cite{boltz2} structures and affinity predictions were generated for the
target metals. For each sequence, we first generated a protein--ligand structure,
identified contacting residues, and then obtained affinity predictions for each of
four candidate binding sites. Metrics were averaged over these four predictions
unless stated otherwise.

Predicted TM-score (pTM) estimates expected global structural similarity using
the predicted aligned-error distribution and the length-normalized TM-score
formulation \cite{tmscore,alphafold}. Values range from zero to one, with larger
values indicating greater confidence in the overall fold and relative domain
arrangement.

Ligand ipTM is a protein--ligand-specific interface predicted TM-score derived
from the AlphaFold-Multimer formulation \cite{afmultimer}. It restricts the
pTM calculation to protein--ligand token pairs. Larger values indicate greater
confidence in the global pose, but do not establish binding or affinity.

Complex ipLDDT is an interface-weighted aggregation of predicted local distance
difference test scores that emphasizes ligand atoms and residues at inter-chain
interfaces. It estimates confidence in local interface geometry rather than
global relative placement.

Complex iPDE summarizes predicted error in distances across inter-chain
contacts. It follows the predicted distance-error formulation used in
AlphaFold~3 \cite{alphafold3}; lower values indicate more confident interface
geometry.

The Boltz-2 confidence score is a composite structure-ranking metric,
\(0.8\times\text{complex pLDDT}+0.2\times\text{ipTM}\), with pTM substituted
for ipTM in single-chain predictions \cite{boltz2software}. Complex pLDDT here is
the mean pLDDT over all tokens of the predicted complex, and is distinct from the
interface-weighted Complex ipLDDT above. It is not a binding
probability or an affinity estimate.

\subsection{Boltz-2 Structural Affinity Metrics}

The Boltz-2 continuous affinity output is expressed as
\(\log_{10}(\mathrm{IC}_{50})\), with \(\mathrm{IC}_{50}\) measured in
\(\mu\mathrm{M}\) \cite{boltz2software}. Lower raw values therefore indicate
stronger predicted affinity. To express this output on a
higher-is-stronger logarithmic scale, we calculated an approximate $\mathrm{pIC}_{50}$ as
\[
\mathrm{pIC}_{50}
=
6-\bigl(\text{continuous affinity output}\bigr).
\]
The maximum approximate $\mathrm{pIC}_{50}$ across the four candidate-site predictions was
retained. This value is a model-derived estimate of $\mathrm{IC}_{50}$ and should not be
interpreted as a direct prediction of \pkd{}.

The Boltz-2 binary affinity probability estimates the probability that the
ligand is a binder rather than a decoy \cite{boltz2software}. Values range from
zero to one, with larger values indicating greater predicted binder likelihood.
This output is trained using different supervision from the continuous affinity
estimate and therefore represents binding-classification confidence rather than
affinity magnitude. The binary affinity probability was averaged across the
four candidate-site predictions.

Pocket-contact count is the number of protein residues assigned to the predicted
ligand-contact region for each candidate site. It describes the size of the
contact-defined pocket rather than structural confidence or binding strength.
Pocket-contact counts were averaged across the four candidate sites.

Because the Boltz-2 affinity heads were developed for protein--small-molecule
interactions, their application to metal ions was treated as an exploratory
structural diagnostic rather than a calibrated estimate of metal-binding
thermodynamics.

\subsection{Auxiliary Structure-Derived Metrics}

We additionally evaluated four auxiliary structure-derived metrics.

Pocket pLDDT is the mean pLDDT of the protein residues assigned to the candidate
binding site and measures local confidence in the predicted pocket geometry.

Pocket-minus-protein pLDDT is the difference between pocket pLDDT and mean
protein pLDDT; positive values indicate that the pocket is predicted more
confidently than the protein overall.

Ligand pLDDT is the mean pLDDT of the
ligand tokens and estimates local confidence in the predicted metal position.
Pocket--ligand PAE is the mean predicted aligned error between the ligand and
the pocket residues; lower values indicate greater confidence in their relative
placement. These quantities are derived from predicted structures and
confidence matrices and are not direct estimates of binding affinity.

\subsection{Boltz-2 Structural Metrics Regression}

To determine whether Boltz-2-derived metrics could be combined to predict binding, we trained two pooled L2-regularized models across the Function-Selected, Co-Generation-Selected, and Structure-Confidence-Selected design panels: a logistic-regression model for terbium binding classification and a ridge-regression model for
continuous Tb and Dy \pkd{} prediction. Both analyses included proteins with mean expression of at least 20 µM and used prespecified
Boltz-2 confidence, affinity, and pocket-contact metrics. Features were aggregated across the four candidate-site predictions, using the mean across sites
except for approximate $\mathrm{pIC}_{50}$, for which the maximum was used. Design-panel identity was not included as a predictor.

For classification, measured Tb \pkd{} $>5.52$ defined binders; lower, censored, or unavailable measurements were assigned to the non-binding class. Cross-validation folds were stratified by design panel and binding class. The continuous model was restricted to proteins with exact, within-range measurements for both Tb and Dy, included a metal indicator, and kept both observations from each protein in the same fold. Both models used 5-fold cross-validation repeated 20 times, with imputation and standardization fitted only within each training fold. The logistic model generated out-of-fold binding probabilities evaluated by ROC AUC, whereas the ridge model generated out-of-fold \pkd{} predictions evaluated by Spearman correlation. Performance was summarized separately for each design panel by averaging across the 20 repeats.

\subsection{Protein Function Measurements}

After the design campaigns were completed, proteins were purchased as DNA, with coding sequences optimized using DNA
Chisel \cite{dnachisel}, and expressed with an in-house cell-free method. Proteins were purified, quantified, normalized by molarity, and
dispensed into 96- and 384-well plates containing pH-controlled buffer. Metal
was titrated into each protein preparation, and a readout of metal--protein
complex formation was collected after each addition. The resulting titration
series was fit with a quadratic binding curve to estimate an apparent
dissociation constant. Positive- and negative-control proteins were screened
with every experiment to monitor execution quality and consistency.

Training data accumulated over successive design--build--test--learn cycles in
which the model proposed panels intended to expand the observed frontier of Tb
and Dy selectivity. Nd/Pr selectivity was also used for data generation but is
outside the principal scope of this study. When available, public proteins,
including lanmodulin orthologs \cite{lanmodulinfamily}, and custom rational designs,
were added to the panels. For the model that designed the proteins in this study, lanmodulin orthologs were not available in the training or testing sets.
Binding constants measured on the same platform supplied the labels for each
protein--metal complex.

\section{Results}

\subsection{Design Panels and Binding Measurement}

We measured four groups of proteins: three from the design strategies described above, and one comprising a large set of wild-type proteins. Each design method generated approximately 100 candidates: Structure-Confidence-Selected produced 93, Co-Generation-Selected produced 104, and Function-Selected produced 100.

All of the proteins aside from the orthologs were passed through Boltz-2 for structure metric scoring; the per-metric results are summarized in Figure~\ref{fig:metric-ranking} and tabulated in Appendix~\ref{app:tables}. We set the binding threshold at \pkd{} $=5.52$, a dissociation constant of 3.0 µM; any measurement weaker than that was beyond the determinable range of the assay and was considered not binding. Constants at or below it are considered to be binding. Proteins that failed to express to the 20 µM needed for the assay were removed and not considered non-binding. This left 96, 76, and 90 scored proteins in the Function-Selected, Co-Generation-Selected, and Structure-Confidence-Selected panels, respectively.

\subsection{No Structural-Confidence Metric Ranks Function Within a Panel}

A selection filter is asked a ranking question: given a set of candidates, does the metric place the ones that will work above the ones that will not? We put that question to each panel separately. Scoring within a panel holds the generative method fixed, so what is left is the signal itself.

\begin{figure*}[t]
  \centering
  \includegraphics[width=0.96\textwidth]{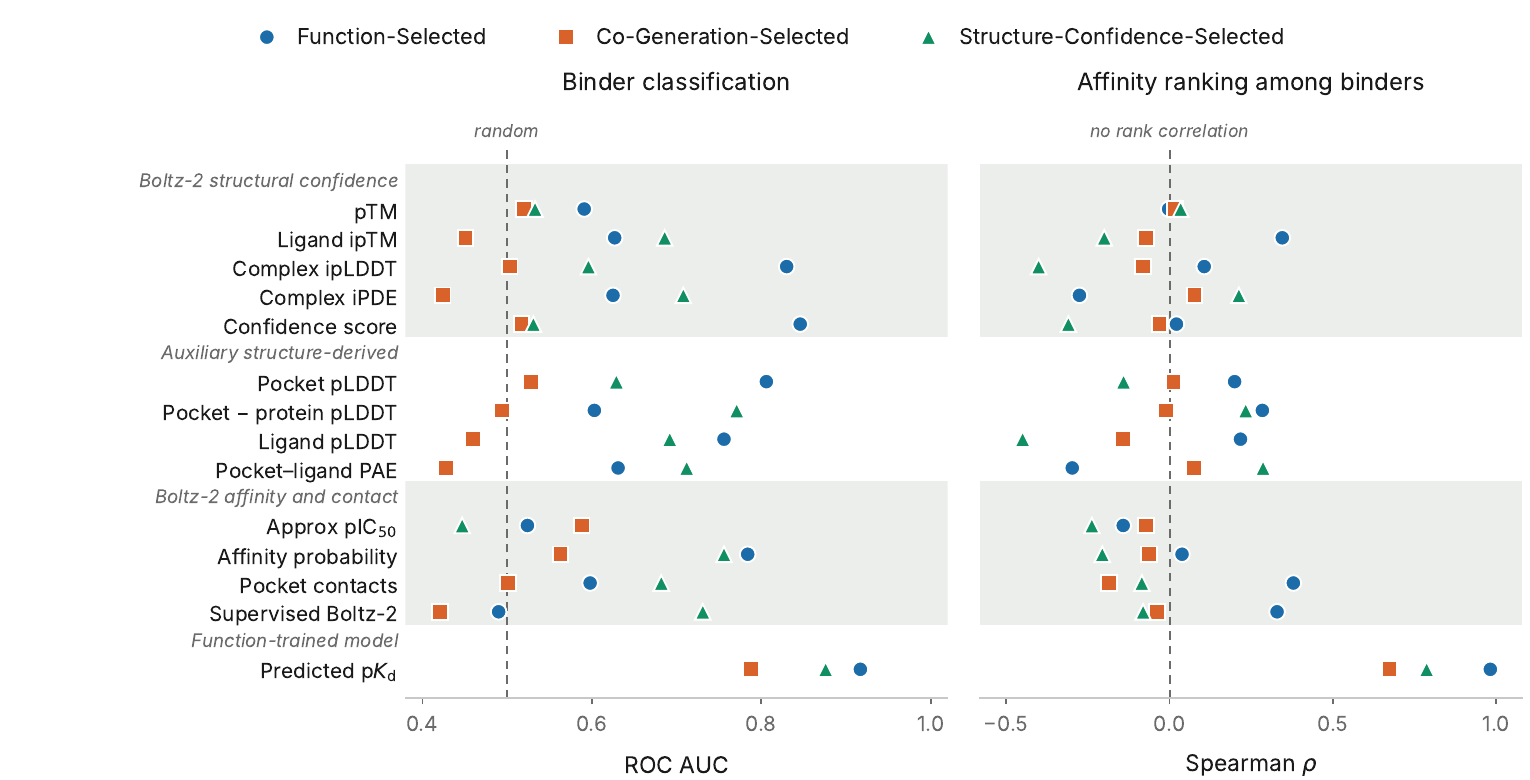}
  \caption{\textbf{No structural-confidence metric ranks function consistently; the
  function-trained prediction does.} Left: ROC AUC for separating binders from
  non-binders within each design panel, against a random-classifier reference of 0.5.
  Right: Spearman rank correlation with measured \pkd{} among binders, against a
  no-correlation reference of 0. Each metric carries one marker per design panel.
  Complex iPDE and pocket--ligand PAE are sign-inverted in the left panel so that
  larger is better throughout. The supervised Boltz-2 baseline is the pooled
  logistic model in the left panel and the pooled ridge model in the right.
  Underlying values are in
  Tables~\ref{tab:roc-auc-by-design-set} and~\ref{tab:spearman-by-design-set}.}
  \label{fig:metric-ranking}
\end{figure*}

The left panel of Figure~\ref{fig:metric-ranking} shows receiver operating characteristic (ROC) area under the curve (AUC) values, computed to measure the quality of these metrics as classification signals for the binding/non-binding label, with the underlying values in Table~\ref{tab:roc-auc-by-design-set}. These metrics include all of the tested proteins, including proteins that showed no binding.  An AUC of 0.5 corresponds to random prediction; $\mathrm{AUC}<0.5$ indicates that low values are predictive of the class, and $\mathrm{AUC}>0.5$ that high values are.

No structural-confidence metric cleared this bar in all three panels. Some reach a fairly strong AUC, but only in method-specific populations: Ligand ipTM and Complex iPDE both fall below 0.5 in the Co-Generation-Selected panel, meaning the metric ranks non-binders above binders there, while reaching 0.69 and 0.71 in the Structure-Confidence-Selected panel. The composite confidence score reaches 0.85 in the Function-Selected panel and is near chance in the other two.

Across all design groups, the function-trained model's predicted \pkd{} had the highest AUC, 0.79 to 0.92, and its rank correlation with measured \pkd{} among binders (Figure~\ref{fig:metric-ranking}, right; Table~\ref{tab:spearman-by-design-set}) was the only one that held across all three panels. The logistic model built from the pooled Boltz-2 metrics reached a comparatively high AUC for the Structure-Confidence-Selected group; it performed much worse on the other two. The Structure-Confidence-Selected panel contributes only 15 binders, so its rank correlations in Table~\ref{tab:spearman-by-design-set} are correspondingly underpowered.

The class means corroborate this. Table~\ref{tab:boltz-metric-means-by-binding-class} gives the means by binding class for each design group. According to the confidence metrics, the Structure-Confidence-Selected designs have the highest pTM, Complex ipLDDT, and composite confidence scores, and Function-Selected proteins had the weakest of the confidence metrics. Ligand ipTM and Complex iPDE do not separate the panels cleanly: all six Ligand ipTM means fall within 0.01 of each other, and Complex iPDE favors Co-Generation-Selected among non-binders but Structure-Confidence-Selected among binders.
The mean Boltz-2 affinity predictions behave the same way, separating the binding and non-binding classes only marginally; what variation they do show tracks design method rather than measured outcome.
The function-trained model's predicted \pkd{} means show a much stronger signal for distinguishing binding from non-binding proteins, with an average difference approaching 1 \pkd{} unit, a 10-fold difference in dissociation constant.

\subsection{Selection on Structural Confidence Produced the Fewest Working Proteins}

A signal that does not rank function within a panel should not improve a panel when it is used to select one. The screening outcome is what that costs in practice.

\begin{figure*}[t]
  \centering
  \includegraphics[width=0.96\textwidth]{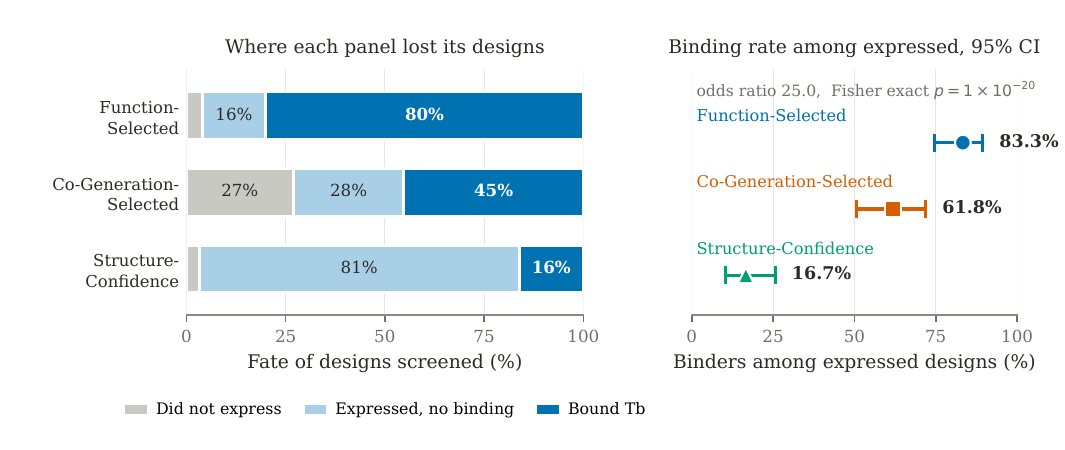}
  \caption{\textbf{Selection on structural confidence produced the fewest working
  proteins.} Left: the fate of every design screened, as a share of that panel's
  total. The Structure-Confidence-Selected panel expressed as well as any other
  and then failed at the binding step, leaving 81\% of its designs expressed but
  inactive; the Co-Generation-Selected panel instead lost 27\% of its designs to
  expression failure. Right: binding rate among expressed designs with 95\% Wilson
  intervals; counts are 80/96, 47/76 and 15/90.}
  \label{fig:binding-outcome}
\end{figure*}

Of the 93 Structure-Confidence-Selected designs, 90 expressed, but only 15 showed binding, 16.7\% (95\% CI 10.4 to 25.7). Of the 104 Co-Generation-Selected designs, 76 expressed and 47 showed binding, 61.8\% (50.6 to 71.9). Of the 100 Function-Selected designs, 96 expressed and 80 showed binding, 83.3\% (74.6 to 89.5). Confidence intervals are Wilson intervals on the expressing population.
Figure~\ref{fig:binding-outcome} summarizes these outcomes.

The three panels do not share a common binding rate ($\chi^{2} = 85.7$, $\mathrm{df} = 2$, $p = 2.5\times10^{-19}$). The gap between the function-guided and structure-confidence panels is the largest contrast in the study: an odds ratio of 25.0 (Fisher exact $p = 1\times10^{-20}$). Function-Selected also exceeded Co-Generation-Selected (odds ratio 3.1, $p = 1.7\times10^{-3}$), and Co-Generation-Selected exceeded Structure-Confidence-Selected (odds ratio 8.1, $p = 2.9\times10^{-9}$). These binding-rate comparisons were specified after the measurements were collected and are reported without multiplicity correction, unlike the prespecified selectivity family in Table~\ref{tab:figure-2-selectivity-statistics}.

The Structure-Confidence-Selected panel therefore had the lowest frequency of active proteins by a large margin despite a 97\% expression rate. These designs folded and expressed as well as any other panel, and then failed at the one property they had been selected for.

\subsection{Selectivity Against the Natural Ortholog Range}

The design objective in this campaign was Tb/Dy selectivity, so we also asked whether any panel moved beyond the selectivity already present in the natural lanmodulin family. This is a narrower question than the binding-rate result above, and the effect sizes are correspondingly smaller.

We performed a two-sided Mann–Whitney comparison against the lanmodulin orthologs to ask if any of these approaches produced proteins that significantly deviate from the existing functional diversity found in nature. The lanmodulin orthologs had a median selectivity of 1.26, with wild-type lanmodulin itself near 1.4. The Function-Selected designs had a median selectivity of 1.70, above the orthologs with a large effect (BH-adjusted $p = 5.6\times10^{-16}$, $r_{\mathrm{rb}} = +0.678$). Expressed as a probability of superiority, a function-guided design outranks a randomly drawn ortholog in 83.9\% of pairs. The Co-Generation-Selected designs, median 1.38, also exceeded the orthologs but with a small effect (BH-adjusted $p = 0.024$, $r_{\mathrm{rb}} = +0.240$), and the Structure-Confidence-Selected designs, median 1.42, were not distinguishable from them (BH-adjusted $p = 0.251$). Both structure-guided panels sat significantly below Function-Selected, at $p = 2.7\times10^{-5}$ and $p = 0.002$ respectively. See Table~\ref{tab:figure-2-selectivity-statistics}.

Table~\ref{tab:figure-2-selectivity-statistics} requires both Tb and Dy measurements to resolve successfully so that a selectivity ratio can be computed, which is why its counts fall below the binder counts in Table~\ref{tab:boltz-metric-means-by-binding-class}. Its cohort assignments and expression values are taken from the design proposal records and the binding dataset, whereas Table~\ref{tab:spearman-by-design-set} takes cohort assignments from the design registry and expression from the expression dataset, and additionally requires Boltz-2 predictions to be available. The two populations therefore differ by a small number of designs.
Such failures are rare and arise from several hard-to-trace sources in the assay; they are treated as random, for example a failed robotic dispense or other mechanical anomaly.

\begin{table*}[t]
  \centering
  \caption{Measured Tb/Dy selectivity and prespecified two-sided Mann--Whitney comparisons. Adjusted p values use Benjamini--Hochberg correction across the full prespecified testing family. Signed rank-biserial r is oriented so positive values indicate higher selectivity for the row cohort relative to the named reference. Bold adjusted p values are significant at 0.05.}
  \label{tab:figure-2-selectivity-statistics}
  \setlength{\tabcolsep}{3.5pt}
  \renewcommand{\arraystretch}{1.12}
  \resizebox{\textwidth}{!}{%
    \begin{tabular}{lrr rrr rrr}
      \toprule
      & \multicolumn{2}{c}{Selectivity} & \multicolumn{3}{c}{vs. Function-Selected} & \multicolumn{3}{c}{vs. Lanmodulin Orthologs} \\
      \cmidrule(lr){2-3} \cmidrule(lr){4-6} \cmidrule(lr){7-9}
      Model/cohort & n & Median  & U & BH $p$ & $r_{\mathrm{rb}}$ (magnitude) & U & BH $p$ & $r_{\mathrm{rb}}$ (magnitude) \\
      \midrule
      Function-Selected & 76 & \textbf{1.70} & \textemdash{} & \textemdash{} & \textemdash{} & 9500.0 & \textbf{\boldmath$5.6\times10^{-16}$} & +0.678 (large) \\
      Co-Generation-Selected & 40 & 1.38 & 770.0 & \textbf{\boldmath$2.7\times10^{-5}$} & $-0.493$ (medium) & 3695.0 & \textbf{0.024} & +0.240 (small) \\
      Structure-Confidence-Selected & 15 & 1.42 & 263.0 & \textbf{0.002} & $-0.539$ (large) & 1319.0 & 0.251 & +0.180 (small) \\
      Lanmodulin Orthologs & 149 & 1.26 & \textemdash{} & \textemdash{} & \textemdash{} & \textemdash{} & \textemdash{} & \textemdash{} \\
      \bottomrule
    \end{tabular}%
  }
\end{table*}

\section{Conclusion}

Structural confidence is a plausibility signal, not a selection criterion for
function. In this bounded rare-earth-binding comparison, the function-guided
designs occupied a more favorable measured Tb/Dy selectivity distribution than
the evaluated reference populations and bound at five times the rate of the
panel selected on structural confidence, while no structure-confidence metric
ranked binders above non-binders consistently within any panel.

The pattern across panels is that confidence metrics can carry some predictive signal, but not consistently: neither across design sets nor across metrics within a set.

The panel with the highest confidence scores expressed at a high rate but scored lowest on function, with the majority of its proteins showing no observable binding.
Conversely, the design set with the lowest overall confidence scores not only showed similarly high expression, but also had the highest rate of observed binding and the most favorable distribution of selectivity, which was the overall design goal.

This comparison is deliberately asymmetric, and the asymmetry is the finding rather than a caveat on it. Structure-confidence metrics are used as design filters precisely because they promise to rank candidates without requiring measurement; that promise is the entire reason pLDDT and related scores appear in design pipelines at all. The function-trained model makes no such promise. It exists because the measurements exist, and it required roughly 2,500 of them. The question this study puts to the two approaches is whether the zero-shot promise holds at the point of selection, and it does not. The cost of the measurement program is what buys the difference.

This result does not make structural modeling unnecessary. It shows that function-guided modeling is the preferable choice when an accurate structure-function rationale is unavailable. The intrinsic disorder of this protein makes a function-specific structural rationale particularly difficult to formulate, and that difficulty is a core bottleneck in structure-based protein design.
Measured function and direct supervision therefore become the natural next step: models that learn function directly from data remove the need to bridge structure to function with hand-built explanations.

Two limits bound the claim. The comparison holds one protein family fixed: the function model was trained largely on lanmodulin variants and evaluated on lanmodulin variants, so what is demonstrated is that in-domain function data outranks zero-shot structural confidence in domain. Whether a function-trained model transfers to an unrelated fold is a separate question that this study does not address. Second, the three panels differ in generative method as well as in selection signal, so part of the gap in binding rate reflects how far each method travels from a scaffold already known to work, independent of which signal performed the selection.

\subsubsection*{Data and Code Availability}

The measurement data and analysis code supporting this study are proprietary to
Aether Biomachines and are not publicly available.

\subsubsection*{Use of Generative AI}

Generative AI language tools were used in the preparation of this manuscript in
three ways: for LaTeX formatting, for coding assistance in the analysis, and for
copy-editing and review of drafts for grammar, internal consistency, and clarity.
The study design, experimental work, data generation, analysis decisions, and all
scientific claims are the authors' own. The authors reviewed and verified the
full manuscript, including every reference, and take full responsibility for its
contents.

\subsubsection*{Acknowledgements}

The authors thank the rest of the Aether team.

\clearpage
\onecolumn
\appendix
\section{Full Metric Tables}
\label{app:tables}

The tables in this appendix give the per-metric values summarized in Figure~\ref{fig:metric-ranking}.

\begin{table}[H]
  \centering
  \caption{Mean raw Boltz-2 metric and predicted-\pkd{} values by measured binding class and protein-design set. Proteins that did not express are counted in the non-binding column of this table, so the non-binding counts here exceed those implied by Table~\ref{tab:roc-auc-by-design-set}, which excludes non-expressing proteins from the ROC populations. The auxiliary structure-derived metrics are reported in Tables~\ref{tab:roc-auc-by-design-set} and~\ref{tab:spearman-by-design-set} only.}
  \label{tab:boltz-metric-means-by-binding-class}
  \vspace{\baselineskip}
  \setlength{\tabcolsep}{3.5pt}
  \renewcommand{\arraystretch}{1.12}
    \resizebox{\textwidth}{!}{%
    \begin{tabular}{lrrrrrr}
      \toprule
      & \multicolumn{2}{c}{Function-Selected} & \multicolumn{2}{c}{Co-Generation-Selected} & \multicolumn{2}{c}{Structure-Confidence-Selected} \\
      \cmidrule(lr){2-3} \cmidrule(lr){4-5} \cmidrule(lr){6-7}
      Metric & Non-binding & Binding & Non-binding & Binding & Non-binding & Binding \\
      \midrule
      Total proteins screened & \multicolumn{2}{c}{100} & \multicolumn{2}{c}{104} & \multicolumn{2}{c}{93} \\
      Proteins successfully expressed & \multicolumn{2}{c}{96} & \multicolumn{2}{c}{76} & \multicolumn{2}{c}{90} \\
      Proteins in binding class & 20 & 80 & 57 & 47 & 78 & 15 \\
      \addlinespace
      \multicolumn{7}{l}{\textit{Boltz-2 structural-confidence metrics}} \\
      pTM & 0.859 & 0.866 & 0.880 & 0.883 & 0.904 & 0.906 \\
      Ligand ipTM & 0.959 & 0.965 & 0.968 & 0.967 & 0.961 & 0.967 \\
      Complex ipLDDT & 0.816 & 0.844 & 0.846 & 0.848 & 0.909 & 0.913 \\
      Complex iPDE (Å; lower is better) & 0.564 & 0.530 & 0.483 & 0.493 & 0.534 & 0.474 \\
      Confidence score & 0.805 & 0.828 & 0.826 & 0.828 & 0.901 & 0.901 \\
      \addlinespace
      \multicolumn{7}{l}{\textit{Boltz-2 affinity and contact metrics}} \\
      Approx $\mathrm{pIC}_{50}$ & 7.232 & 7.264 & 7.208 & 7.246 & 7.344 & 7.332 \\
      Affinity probability binary mean & 0.727 & 0.733 & 0.733 & 0.732 & 0.732 & 0.738 \\
      Pocket contacts & 8.312 & 8.387 & 8.470 & 8.449 & 7.777 & 8.267 \\
      \addlinespace
      \multicolumn{7}{l}{\textit{Predicted \pkd{}}} \\
      Function-model predicted \pkd{} & 5.724 & 6.836 & 5.736 & 6.375 & 5.706 & 7.225 \\
      Supervised Boltz-2 (ridge) & 6.607 & 6.707 & 6.833 & 6.831 & 7.015 & 7.049 \\
      \bottomrule
    \end{tabular}%
  }
  \vspace{\baselineskip}
\end{table}

\begin{table}[H]
  \centering
  \caption{ROC area under the curve by protein-design set. Plotted in the left panel of Figure~\ref{fig:metric-ranking}.}
  \label{tab:roc-auc-by-design-set}
  \vspace{\baselineskip}
  \setlength{\tabcolsep}{8pt}
  \renewcommand{\arraystretch}{1.12}
  \resizebox{\textwidth}{!}{%
  \begin{tabular}{lccc}
    \toprule
    Metric & Function-Selected & Co-Generation-Selected & Structure-Confidence-Selected \\
    \midrule
    Total proteins & 100 & 104 & 93 \\
    Expressing proteins eligible for ROC & 96 & 76 & 90 \\
    Binders & 80 & 47 & 15 \\
    Non-expressing proteins (excluded) & 4 & 28 & 3 \\
    \addlinespace
    \multicolumn{4}{l}{\textit{Boltz-2 structural-confidence metrics}} \\
      pTM & 0.591 & 0.520 & 0.533 \\
      Ligand ipTM & 0.627 & 0.451 & 0.686 \\
      Complex ipLDDT & 0.830 & 0.503 & 0.596 \\
      Complex iPDE (Å; sign-inverted for ROC) & 0.625 & 0.424 & 0.708 \\
      Confidence score & 0.846 & 0.517 & 0.531 \\
    \addlinespace
    \multicolumn{4}{l}{\textit{Auxiliary structure-derived metrics}} \\
      Pocket pLDDT & 0.806 & 0.528 & 0.629 \\
      Pocket-minus-protein pLDDT & 0.603 & 0.494 & 0.771 \\
      Ligand pLDDT & 0.756 & 0.460 & 0.692 \\
      Pocket--ligand PAE (Å; sign-inverted for ROC) & 0.631 & 0.428 & 0.712 \\
    \addlinespace
    \multicolumn{4}{l}{\textit{Boltz-2 affinity and contact metrics}} \\
      Approx $\mathrm{pIC}_{50}$ & 0.524 & 0.588 & 0.447 \\
      Affinity probability binary mean & 0.784 & 0.563 & 0.756 \\
      Pocket contacts & 0.598 & 0.501 & 0.682 \\
      Supervised Boltz-2 (logistic) & 0.490 & 0.421 & 0.731 \\
    \addlinespace
    \multicolumn{4}{l}{\textit{Function-model-predicted \pkd{}}} \\
      Predicted \pkd{} & 0.917 & 0.788 & 0.876 \\
    \bottomrule
  \end{tabular}
  }
  \vspace{\baselineskip}
\end{table}

\begin{table}[H]
  \centering
  \caption{Spearman rank correlation ($\rho$) by protein-design set, over designs with Tb and Dy measurements within assay range and Boltz-2 predictions available at the time of analysis. Plotted in the right panel of Figure~\ref{fig:metric-ranking}.}
  \label{tab:spearman-by-design-set}
  \vspace{\baselineskip}
  \setlength{\tabcolsep}{8pt}
  \renewcommand{\arraystretch}{1.12}
    \resizebox{\textwidth}{!}{%
  \begin{tabular}{lccc}
    \toprule
    Metric & Function-Selected & Co-Generation-Selected & Structure-Confidence-Selected \\
    \midrule
    Measured Binders & 75 & 42 & 15 \\
    \addlinespace
    \multicolumn{4}{l}{\textit{Boltz-2 structural-confidence metrics}} \\
      pTM & $-$0.003 & 0.017 & 0.034 \\
      Ligand ipTM & 0.345 & $-$0.073 & $-$0.200 \\
      Complex ipLDDT & 0.106 & $-$0.082 & $-$0.401 \\
      Complex iPDE (Å; lower is better) & $-$0.276 & 0.076 & 0.212 \\
      Confidence score & 0.021 & $-$0.031 & $-$0.310 \\
    \addlinespace
    \multicolumn{4}{l}{\textit{Auxiliary structure-derived metrics}} \\
      Pocket pLDDT & 0.199 & 0.012 & $-$0.141 \\
      Pocket-minus-protein pLDDT & 0.284 & $-$0.012 & 0.233 \\
      Ligand pLDDT & 0.217 & $-$0.142 & $-$0.450 \\
      Pocket--ligand PAE (Å; lower is better) & $-$0.298 & 0.075 & 0.286 \\
    \addlinespace
    \multicolumn{4}{l}{\textit{Boltz-2 affinity and contact metrics}} \\
      Approx $\mathrm{pIC}_{50}$ & $-$0.142 & $-$0.073 & $-$0.238 \\
      Affinity probability binary mean & 0.038 & $-$0.063 & $-$0.206 \\
      Pocket contacts & 0.379 & $-$0.185 & $-$0.085 \\
      Supervised Boltz-2 (ridge) & 0.329 & $-$0.039 & $-$0.081 \\
    \addlinespace
    \multicolumn{4}{l}{\textit{Function-model-predicted \pkd{}}} \\
      Predicted \pkd{} & 0.982 & 0.673 & 0.787 \\
    \bottomrule
  \end{tabular}
  }
  \vspace{\baselineskip}
\end{table}

\end{document}